\documentclass[aps,prb,groupedaddress,twocolumn,showpacs,superscriptaddress,amssymb,amsmath,notitlepage]{revtex4-1}

\usepackage{algorithm}
\usepackage{algpseudocode}

\usepackage{amsmath}
\usepackage{physics}
\usepackage{verbatim}
\usepackage{graphicx}
\usepackage{amssymb}
\usepackage{amsfonts}
\usepackage{color}
\usepackage{bm}        
\usepackage{comment}
\usepackage{placeins}
\usepackage[mathscr]{eucal}
\usepackage[colorlinks, linkcolor=red,citecolor=blue,urlcolor=blue]{hyperref}
\usepackage[normalem]{ulem}
\usepackage[all]{hypcap}
\usepackage{multirow}
\usepackage[dvipsnames,usenames,table]{xcolor}
\usepackage{dcolumn}
\usepackage[capitalise]{cleveref}
\usepackage{hyperref}
\usepackage{bm}
\usepackage{epsf}
\usepackage{subfigure}
\usepackage{amsfonts}
\usepackage{epstopdf}%

\usepackage{environ}
\usepackage{pifont}
\usepackage{xcolor}

\makeatletter
\newsavebox{\measure@tikzpicture}
\NewEnviron{scaletikzpicturetowidth}[1]{%
  \def\tikz@width{#1}%
  \begin{lrbox}{\measure@tikzpicture}%
  \BODY
  \end{lrbox}%
  \pgfmathparse{#1/\wd\measure@tikzpicture}%
  \BODY
}

\newcommand{\be}{\begin{equation}}
\newcommand{\ee}{\end{equation}}
\newcommand{\bea}{\begin{eqnarray}}
\newcommand{\eea}{\end{eqnarray}}

\definecolor{mblue}{rgb}{0.0,0.45,0.74}
\usepackage{hyperref}
\usepackage{braket}
\usepackage{amsfonts}
\usepackage{tikz}
\usepackage{pgfplots}
\pgfplotsset{compat=1.15}
\usepackage{mathrsfs}
\usetikzlibrary{arrows}

\allowdisplaybreaks
\begin{document}
\title{Reply to ``Comment on ``Unified Framework for Open Quantum Dynamics with Memory''''}
\author{Felix Ivander}
\affiliation{Quantum Science and Engineering, Harvard University, Cambridge, MA, USA}
\author{Lachlan P. Lindoy}
\affiliation{National Physical Laboratory, Teddington, TW11 0LW, United Kingdom}
\author{Joonho Lee}
\email{joonholee@g.harvard.edu}
\affiliation{Department of Chemistry and Chemical Biology, Harvard University, Cambridge, MA, USA}
\affiliation{Google Quantum AI, Venice, CA, USA}

\begin{abstract}
    We present our response to the commentary piece by Makri {\it et al.} [arXiv:2410.08239], which raises critiques of our work [Nat. Commun. 15, 8087 (2024)]. In our paper, we considered various settings of open-quantum system dynamics, including non-commuting, non-diagonalizable system-bath coupling, and bosonic/spin/fermionic baths. For these, we showed a direct and explicit relationship between the discrete-time memory kernel ($\mathcal K$) of the generalized quantum master equation (GQME) and the discrete-time influence functions ($I$) of the path integrals. As an application of this, we showed one can construct $\mathcal K$ without projection-free dynamics inputs that conventional methods require, and we also presented a quantum sensing protocol that characterizes the bath spectral density from reduced system dynamics.  As the Comment focused on the relationship between ($\mathcal K$) and $I$ in one specific setup (i.e., commuting, diagonalizable system-bath coupling with a bosonic bath), we focus on that aspect in this response. In summary, we could not find a set of equations that explicitly connect $I$ and $\mathcal K$ from Makri's 2020 paper [J. Chem. Theory Comput. 16, 4038 (2020)]. Furthermore, while our analysis is specific to the choice of discretization of path-integral and GQME, we have not found issues with the GQME discretization employed. As per critiques on citations, in our paper, we note that we had acknowledged 
    Makri's driven SMatPI work and Wang and Cai's tree-based SMatPI work for the number of Dyck paths needed for the computation of the memory kernel.
\end{abstract}

\maketitle

In our recent paper\cite{ivander2024unified}, we established the {\it explicit} relationship between the Nakajima-Zwanzig memory kernel ($\mathcal{K}$) and the influence functions ($I$) for a wide range of open quantum dynamics setups. We then presented several applications that were made possible based on this explicit connection. Before responding to the recent Comment\cite{makri2024commentunifiedframeworkopen} by Makri {\it et al.}, we summarize our main contributions as follows:
\begin{enumerate}
    \item We provided an explicit connection between discrete-time influence functions and memory kernels for open quantum system dynamics with bilinear system-bath couplings and harmonic baths. By the ``explicit'' connection, we mean that we provide a set of equations that express $\mathcal K$ in terms of $I$ and vice versa (see Eqns. 7-10 and Eqns. 12-14 in Ref.\citenum{ivander2024unified}.) For the time-independent Hamiltonian, we made such connections for these classes of setups: 
    \begin{enumerate}
        \item[(a)] Class 1: All system terms ($\hat{S}_{i,\alpha}$) in the system-bath interactions ($\hat{H}_I = \sum_{i,\alpha} \hat{S}_{i,\alpha}\otimes\hat{B}_{i,\alpha}$) are all diagonalizable and commute,
        \item[(b)] Class 2: No terms ($\hat{S}_{i,\alpha}$) in the system-bath interactions ($\hat{H}_I$) commute, but each $\hat{S}_{i,\alpha}$ is diagonalizable,
        \item[(c)] Class 3: There are common baths for some coupling terms and their system operators may or may not commute,
        \item[(d)] Class 4: No terms ($\hat{S}_{i,\alpha}$) in the system-bath interactions ($\hat{H}_I$) commute and are diagonalizable,
    \end{enumerate}
    considering various Gaussian baths (bosonic, spin, and fermionic).
    \item This enables the construction of $\mathcal{K}$ without conventional projection-free dynamics inputs.
    \item We showed that it is possible to characterize the bath spectral function, $J(\omega)$, given the tomographically complete reduced system dynamics, $\rho(t)$. Such a procedure was prescribed in full for problems in Class 1 and outlined for other Classes.
\end{enumerate}
Makri {\it et al.} \cite{makri2024commentunifiedframeworkopen} 
presents their critiques only about our contribution 1 (a). 
Class 1 poses the simplest form of a system-bath Hamiltonian, and we presented a more detailed analysis for that case. Similarly, other groups have focused primarily on this case.
Furthermore, our main contributions are not directly related to new numerical approaches but are most meaningful for their explicit analytical derivations. 
With this in mind, we shall focus on our Contribution 1 (a) and related analyses. 
In particular, the authors of the Comment raised the following points:
\begin{enumerate}
    \item The relationship between $\mathcal K$ and $I$ was established in 2020 in Ref. \citenum{makri2020small} by Makri.
    \item The absence of endpoint effects in the memory kernel is due to the crude generalized quantum master equation (GQME) discretization.
    \item The lack of citation to Makri's SMatPI work on driven open quantum system dynamics in Ref. \citenum{DrivenSmatpi}.
    \item The Dyck path section follows the diagrammatic analysis developed by Wang and Cai in Ref. \citenum{tree-smatpi}.
\end{enumerate}
We hope to address these points in this reply.

{\it Point \#1.} Makri {\it et al.} claims that the relationship between $\mathcal K$ and $I$ was established in the work by Makri in 2020~\cite{makri2020small}.
After a close inspection of the cited paper again, we could not find a set of equations in Ref.~\citenum{makri2020small} that write $\mathcal K$ in terms of $I$ along the discretized time axis. 
As pointed out in the Comment, the cited paper has the following sentences:
{\it ``This structure is reminiscent of the Nakajima–Zwanzig generalized quantum master equation (GQME), where the time derivative of the RDM depends on the RDM history through a simple time integral. In fact, eq 4.6 bears a close resemblance to the transfer tensor scheme (TTM), which in the $\Delta t\rightarrow 0$ limit is equivalent to the GQME.''}
TTM is indeed a discretized version of the GQME, as noted in Ref.~\onlinecite{CaoTTM}, and drawing a correspondence between the memory kernel of TTM and $I$ would achieve our Contribution 1(a).
The cited paper only mentions an observation of a close resemblance between SMatPI and TTM and does not explicitly construct $\mathcal K$ in terms of $I$ like our work did. In Ref.~\citenum{makri2020small}, furthermore, Makri also noted differences between SMatPI and TTM based on the endpoint effects, which makes it unclear whether explicit connections were easily accessible in their analysis.

{\it Point \#2.} Our analysis is based on the expansion of the discrete-time system propagator in terms of influence functions, $I$, as is done in SMatPI~\cite{makri2020small} and the earlier work by Golosov, Friesner, and Pechukas~\cite{golosov1999efficient} that used such an analysis to motivate an approximate expression for the memory matrices. One critical difference between our use of this expansion and others' is the Trotter ordering. We place $\exp(-i\hat{H}_S\Delta t/2)$ at the boundaries, while others have placed $\exp(-i\hat{H}_S\Delta t)$ in the middle.
This seemingly innocent Trotter ordering change enables the decomposition of the system propagator $\mathbf{U}$ in terms of fully time-translational tensors, $\tilde{\mathbf U}$.
With the decomposition of $\tilde{\mathbf U}$ in terms of $I$, we relate $I$ to the memory kernel $\mathbf K$ in the discretized GQME.
As pointed out by the Comment, the discretization we employ is correct up to $\mathcal O(\Delta t^3)$ for $\tilde{\mathbf U}$ and up to $\mathcal O(\Delta t^2)$ for $\mathcal K$. Our analysis is valid between these specific discretizations. Makri {\it et al.} refers to this GQME discretization as ``crude'' approximations, but as we show in our paper (see, for instance, Fig. 3)~\cite{ivander2024unified} and also evidenced by TTM,~\cite{CaoTTM} this does not raise significant concerns in numerical examples.
Lastly, Makri {\it et al.} attributes the absence of endpoint effects (i.e., the appearance of time-translationally invariant tensors) to the ``crude'' approximation. In this particular setting, i.e., where $\hat{H}_S$ is time-independent, the memory kernel in the continuous-time limit is time translationally invariant. Therefore, we respectfully disagree that it is formally less desirable to prioritize time translational invariance over reducing the Trotter error. Furthermore, one can derive a time-translationally invariant memory kernel with an arbitrary order time step error using projection operator techniques.

{\it Point \#3.} As an extension of our analysis, we also considered a scenario with explicit time dependence in the system Hamiltonian. While our paper does not propose any numerical methods, we cited Makri's driven SMatPI approach~\cite{DrivenSmatpi} in the Supplementary Notes H of our paper to note differences if one were to develop new methods based on our formulation.
In essence, with our Trotter ordering, it is possible to construct the two-time memory kernel via a tensor contraction between a time-dependent tensor, $P$, and a time-independent influence function tensor, $T$, as explained in our work.
Our approach does not require computing the influence function tensor every time step, unlike Makri's driven extension to SMatPI~\cite{DrivenSmatpi}, which calculates the memory matrices at each time step.
However, our approach requires the storage of high-dimensional tensors (whereas SMatPI works only with smaller tensors) and should probably be combined with tensor network approaches similar to those recently developed.~\cite{TEMPO2018}
Our formulation is more closely related to process tensor methods~\cite{jorgensen2019,Jorgensen2020Nov} where a similar decomposition is possible for the system propagator $U$. 

{\it Point \#4.} It is true that our Dyck path section has similar analyses as those presented by Wang and Cai,~\cite{tree-smatpi} which we cited both in our main text and in the corresponding supplementary section. We only cited their preprint,\cite{tree-smatpi} and not the final version published in the journal.\cite{Wang2024Jan} This was an oversight on our part. Wang and Cai used the Dyck paths to analyze the computational cost of their tree-based SMatPI algorithm. 
Although Wang and Cai only considered the SMatPI kernel matrices (i.e., different from the memory kernel in our paper), we cited their work for the number of Dyck paths in $\mathcal K$, i.e., the Catalan number. Our work also presents a general algorithm to use the Dyck paths directly to obtain terms composing the memory kernels at arbitrary time points, which is different from the tree-based algorithm presented by Wang and Cai.
We also noted that our Dyck path construction is limited to the Class 1 problems. 

{\it Conclusions.} We have presented our perspectives on the critiques presented by Makri {\it et al.} We identified four main critiques from their commentary piece and addressed each point. 
We will summarize these points again. First, we could not find a set of equations that explicitly connect $I$ and $\mathcal K$ from Makri's 2020 paper. Second, while our analysis is specific to the choice of discretization of path-integral and GQME, we have not found issues with our choice of GQME discretization that has an $\mathcal O(\Delta t^2)$ error.
Third, we had cited Makri's driven SMatPI work in Supplementary Notes H, noting main differences. Fourth, we had cited Wang and Cai's tree-based SMatPI work for the number of Dyck paths needed for the computation of the memory kernel.
Last but not least, our work covers other types of baths (spin and fermionic) and problems not in Class 1, with applications to Hamiltonian learning, as highlighted as main contributions in our paper.
We hope that our response clarifies misunderstandings of our work's contribution and relationship to Makri's SMatPI works and the related tree-based SMatPI work of Wang and Cai.

{\it Acknowledgement.} F.I. and J.L. were supported by Harvard University’s startup funds. L.P.L. acknowledges the support of the Engineering and Physical Sciences Research Council [grant EP/Y005090/1]. The authors thank Nathan Ng, David Reichman, and David Limmer for their encouragement and discussion.

\bibliography{refs}

\end{document}